\documentclass[twocolumn, pra, superscriptaddress]{revtex4-1}
\usepackage{amssymb}
\usepackage{amsmath}
\usepackage{epsfig}
\usepackage{graphics, graphicx}
\usepackage{bbold}
\usepackage{psfrag}
\usepackage{mathcomp}
\usepackage{subfigure}
\usepackage{verbatim}
\usepackage{enumitem}
\usepackage[table]{xcolor}
\usepackage{bm}
\usepackage[marginal]{footmisc} 
\usepackage{hyperref}
\usepackage{siunitx}

\def\cp#1{\mathbf{#1}}

\begin{document}

\title{Confinement-induced bound state and effective potential for microwave-shielded ultracold molecules}
\author{Linyi Zhang}
\affiliation{Beijing National Laboratory for Condensed Matter Physics, Institute of Physics, Chinese Academy of Sciences, Beijing, 100190, China}
\affiliation{School of Physical Sciences, University of Chinese Academy of Sciences, Beijing 100049, China}
\author{Xiaoling Cui}
\email{xlcui@iphy.ac.cn}
\affiliation{Beijing National Laboratory for Condensed Matter Physics, Institute of Physics, Chinese Academy of Sciences, Beijing, 100190, China}
\date{\today}

\begin{abstract}
We investigate the bound state and effective potential of microwave-shielded ultracold molecules under longitudinal  confinement, which effectively creates a quasi-two-dimensional (q2D) geometry. We find that  the longitudinal confinement  facilitates  bound-state formation in both bosonic and fermionic molecules, leading to  confinement-induced bound states that are absent in free space. To describe these bound states, we establish an effective 2D potential by incorporating the full interaction effect along the longitudinal direction. Compared with other simplified 2D potentials, the effective potential accurately predicts the binding energies of ultracold molecules over a much broader parameter regime  of microwave coupling and confinement strength. Remarkably, it can even predict  deep bound states with binding energies well exceeding the confinement frequency. By establishing this effective  potential, our work paves the way for efficiently exploring the rich low-dimensional physics of microwave-shielded ultracold molecules.
\end{abstract}
\maketitle

\section{Introduction}

The discovery of {\it confinement-induced resonance} (CIR)\cite{Olshanii1, Olshanii2, Petrov1, Petrov2, Blume, Pricoupenko, Tan} in ultracold atoms has reshaped our understanding of effective low-dimensional scattering, in which external confinement modifies not only the spatial geometry of the atoms but also their low-energy collision properties. For these atomic systems with short-range (contact) interactions, the low-dimensional scattering is fully characterized by a reduced scattering length that depends sensitively on the original three-dimensional (3D) scattering length and the confinement strength; by tuning these parameters, one can engineer both CIR and the confinement-induced bound state (CIBS), as observed in ultracold experiments\cite{s_wave_expt, p_wave_expt, splitted_CIR_expt}. Understanding these confinement-induced scattering properties has laid the foundation for subsequent explorations of intriguing low-dimensional phenomena in ultracold gases\cite{review1, 1D_review, 2D_review}.

Given the success of CIR in atomic systems, its extension to long-range interacting dipolar systems has attracted considerable interest\cite{Santos, Zinner, Schmelcher, Deuretzbacher, Deuretzbacher2, Guan, Shi_Yi}. However, this extension is not straightforward: the anisotropy of the dipolar interaction induces mixed scattering between different partial-wave channels, and more importantly, in the presence of a long-range $1/r^3$ potential, the low-dimensional scattering no longer follows the scattering length description\cite{Guan, long_range, Girardeau}. Indeed, recent experiments on low-dimensional dipolar gases have provided evidence that the dipolar force cannot simply be replaced by a contact interaction\cite{Lev_expt, Jo_expt}, calling for a refined understanding of confinement-induced physics under long-range interactions.

\begin{figure}[t]
\includegraphics[width=8.5cm]{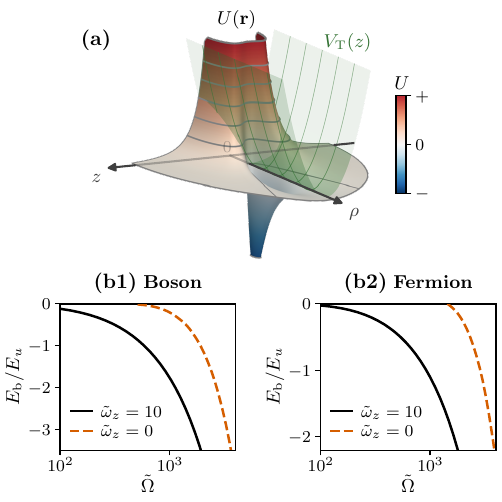}
\caption{(Color online). (a) Schematics of interaction potential $U$ (color-mapped surface) and longitudinal harmonic potential $V_{\rm T}$ (green surface) in the relative coordinate space of two molecules. Here a single $\sigma^{+}$-polarized microwave field is considered, leading to a fully symmetric $U$ in $xy$ plane that solely depends on $\rho\equiv\sqrt{x^2+y^2}$ and $z$. (b1,b2) Ground state binding energies of bosonic (b1) and fermionic (b2) molecules as functions of $\tilde{\Omega}$ with (solid) and without (dashed) longitudinal confinement. Here the energy unit is $E_u$ (see text). }  \label{fig_CIBS}
\end{figure}

In recent years, the rapidly developing field of microwave-shielded ultracold molecules has offered new opportunities  for studying CIR physics under long-range interactions. Under  microwave shielding\cite{Hutson, Quemener, Doyle, Luo1, Luo2, Luo3, Luo4, Wang, Will1, Will2, Will_new,Wang2, ZhangJ}, the inter-molecule potential features an anisotropic long-range tail ($\sim1/r^3$) and a repulsive shielding core ($\sim1/r^6$\cite{Shi1, double1, double2}) that expands over thousands of Bohr radii, which differs strongly from the dipolar potential in magnetic atoms. The presence of the large shielding core  stabilizes the molecular gas from inelastic collision, and has thus facilitated  milestone observations of BECs and self-bound droplets in bosonic NaCs\cite{Will2, Will_new} and NaRb\cite{Wang2} molecules, as well as degenerate gases and tetratomic bound states of fermionic NaK molecules\cite{Luo1, Luo2, Luo3, ZhangJ, Luo4}. In many of these experiments, the molecules experience a tighter confinement along the longitudinal direction than in the transverse directions, effectively creating a quasi-two-dimensional (q2D) configuration\cite{Will2, Will_new, Wang2, Luo1, Luo3,Luo4}. It is therefore pertinent to ask how to describe the system when the longitudinal confinement is sufficiently tight to drive it into the effective 2D regime. Existing theories of 2D polar molecules have relied on simplified planar potentials based on certain assumptions\cite{Zollor1, Zollor2, Gora1, Gora2, DWWang, Shi2, Zillich, Reimann}, whereas a full account of the longitudinal interaction effect has not yet been provided.

In this work, we aim to address the effective 2D scattering of microwave-shielded ultracold molecules under longitudinal confinement. Specifically, we take the example of a single $\sigma^+$-polarized microwave field together with a harmonic confinement along $z$, yielding the potentials illustrated in Fig.~\ref{fig_CIBS}(a). We find that the presence of longitudinal confinement facilitates  bound-state formation in both bosonic and fermionic molecules, giving rise to CIBS as shown in Fig.~\ref{fig_CIBS}(b1,b2). To describe these bound states, we establish an effective 2D potential by incorporating the full interaction effect along the longitudinal direction, which goes well beyond  previous simplified treatments based on the lowest-mode approximation or slice section at the $xy$ plane.  As a result, the effective potential is able to accurately predict the binding energies of bosonic and fermionic molecules over a wide parameter regime of microwave coupling and confinement strength (Fig.~\ref{fig_diagram}). Remarkably, it can even predict deep bound states, with binding energies well exceeding  the confinement frequency. Our results are directly applicable to realistic bosonic and fermionic molecules, which  may hopefully serve as a guideline for future explorations of the rich few- and many-body physics in low-dimensional polar molecules with microwave shielding.

\section{Model} 

We start from the following Hamiltonian that governs the relative motion of two molecules: 
\begin{equation}
H({\cp r})=-\frac{\hbar^2}{2\mu}\nabla_{\cp r}^2+U({\cp r})+V_{\rm T}(z),
\end{equation}
where ${\cp r}=(\bm{\rho}, z)$ is the relative coordinate  and $\mu=m/2$ is the reduced mass; $V_{\rm T}(z)=\mu \omega_z^2z^2/2$ is the harmonic trapping potential along longitudinal direction; $U$ is the interaction potential of polar molecules dressed by a $\sigma^+$-polarized microwave field, as given by\cite{Shi1}  
\begin{equation}
U({\cp r})=\frac{C_3}{r^3} \Big(3\cos^2\theta -1 \Big) + \frac{C_6}{r^6} \Big(1-\cos^4\theta \Big).
\end{equation}
Here $r\equiv |{\cp r}|$ and $\cos\theta= z/r$; $C_3=\frac{d^2}{48\pi\epsilon_0(1+\delta_r^2)}$,  $C_6 = \frac{d^4}{128\pi^2\epsilon_0^2\hbar\Omega(1+\delta_r^2)^{3/2}}$ ($\delta_r = \frac{|\delta|}{\Omega}$), with $\Omega$ and $\delta$ respectively the frequency and detuning of microwave field; $d$ is the dipole moment. In this work we take $\delta_r=0$ and thus an effective dipole moment can be defined as  $d_{\rm eff}=d/\sqrt{12}$\cite{Zwierlein2022}, which gives the dipole length $l_d\equiv \frac{m}{\hbar^2}\frac{d_{\rm eff}^2}{4\pi\epsilon_0}$\cite{Bohn5, Hutson4}. Specifically,  for bosonic NaRb\cite{Wang,Wang2}, NaCs\cite{Will1,Will2,Will_new}, CaF\cite{Doyle} and fermionic NaK\cite{Luo1,Luo2,Luo3, ZhangJ, Luo4} molecules, we have $l_d=1.40,\ 4.38,\ 0.69,\ 0.58 \ (\mu{\rm m})$ respectively. Since $l_d$ is generally much larger than the size of shielding core ($\sim$ a few to tens of nm, which is also the typical size of two-molecule bound state), in this work we take the length unit as $r_u=l_d/10$ to better resolve  bound states. Accordingly,  the energy unit is taken as $E_u=\hbar^2/(mr_u^2)$, which gives $E_u/h=4.68, 0.34, 35.79, 47.70$ (kHz) for NaRb, NaCs, CaF and NaK respectively. When scaled by $E_u$, only two dimensionless parameters remain, namely
\begin{equation*}
\tilde{\Omega}\equiv\frac{\hbar\Omega}{E_u}; \ \ \ \ \ \tilde{\omega}_z\equiv\frac{\hbar\omega_z}{E_u},
\end{equation*} 
which universally characterize the bound state solutions in both bosonic and fermionic molecules.




\section{Exact solutions}

We have exactly solved the bound state of two molecules from the Schr{\"o}dinger equation
\begin{equation}
H({\cp r})\Psi({\cp r})=E\Psi({\cp r}), \label{H}
\end{equation} 
which gives the binding energy $E_{\rm b}=E-\hbar\omega_z/2$. To solve (\ref{H}), we have discretized the ${\cp r}$ space and exactly diagonalized the $H$ matrix with boundary condition $\Psi({\cp 0})=\Psi({\cp R}_c)=0$, with ${\cp R}_c$ the large system boundary. In our numerics, we have adopted different ${\cp R}_c$ and different discretization schemes to ensure the convergence of our results. In Fig.~\ref{fig_CIBS}(b1,b2), we show  $E_{\rm b}$ as a function of $\tilde{\Omega}$ for bosonic and fermionic molecules with a typical longitudinal confinement  $\tilde{\omega}_z$.  In comparison with the free-space bound states ($\tilde{\omega}_z=0$),  these states are more deeply bound (with larger $|E_{\rm b}|$) and can emerge at much smaller $\tilde{\Omega}$.  Therefore, these bound states are closely associated with the q2D geometry and  thus called confinement-induced bound states (CIBS).


\section{Effective 2D theory} 

To understand the physical origin of CIBS, it is necessary to establish an effective 2D model for molecules moving in the confined geometry. Before proceeding, let us first recall two simplified 2D potentials. The first is based on the lowest-mode approximation, giving 
\begin{equation}
U_{\rm LMA}(\bm{\rho})=\int dz  |\phi_0(z)|^2 U({\cp r}), \label{V_LMA}
\end{equation} 
with $\phi_0$ describing the lowest harmonic level of relative motion along $z$. The second is to directly select a slice potential at the $xy$ plane, namely
\begin{equation}
U_{\rm slice}(\bm{\rho})=U(\bm{\rho},z=0). \label{V_slice}
\end{equation} 
Apparently, these potentials rely on certain assumptions that either neglect the longitudinal higher-mode contributions in the virtual scattering processes ($U_{\rm LMA}$) or neglect the quantum fluctuations beyond $xy$ plane ($U_{\rm slice}$). 

To fully incorporate the interaction effect along longitudinal direction, we adopt the adiabatic representation\cite{Blume2,Cavagnero,Greene3} and express the relative wavefunction as 
\begin{equation}
\Psi({\cp r})=\sum_{\nu} F_{\nu}(\bm{\rho})\chi_{\nu}(\bm{\rho},z) \label{adiabatic_wf}
\end{equation} 
where $\chi_{\nu}$ is the $\nu$-th eigenstate and satisfies
\begin{equation}
\left( -\frac{\hbar^2}{2\mu}\partial_z^2+V_{\rm T}(z)+U(\bm{\rho},z) \right) \chi_{\nu}(\bm{\rho},z)=\epsilon_{\nu}(\bm{\rho}) \chi_{\nu}(\bm{\rho},z). \label{eq_z}
\end{equation}
Plugging (\ref{adiabatic_wf}) into (\ref{H}) and neglecting all off-diagonal couplings between different $\nu$, we arrive at 
\begin{equation}
\left( -\frac{\hbar^2}{2\mu}\nabla_{\bm{\rho}}^2+\epsilon_{\nu}(\bm{\rho}) \right) F_{\nu}(\bm{\rho})=E_{\nu}F_{\nu}(\bm{\rho}). \label{H2d}
\end{equation}
The lowest level ($\nu=0$) then corresponds to the ground state of effective 2D model with wavefunction $F_{\nu=0}$, binding energy $E_{{\rm 2D}}\equiv E_{\nu=0}-\hbar\omega_z/2$ and an effective 2D potential 
\begin{equation}
U_{\rm 2D}(\bm{\rho})=\epsilon_{\nu=0}(\bm{\rho})-\hbar\omega_z/2. \label{U2d}
\end{equation}
Note that in obtaining $U_{\rm 2D}$, we have neglected the diagonal correction $\propto \langle\chi_{0}|\nabla_{\bm{\rho}}^2\chi_{0}\rangle_z$, whose effect will be discussed later. 

\begin{figure}[t]
\includegraphics[width=8.5cm]{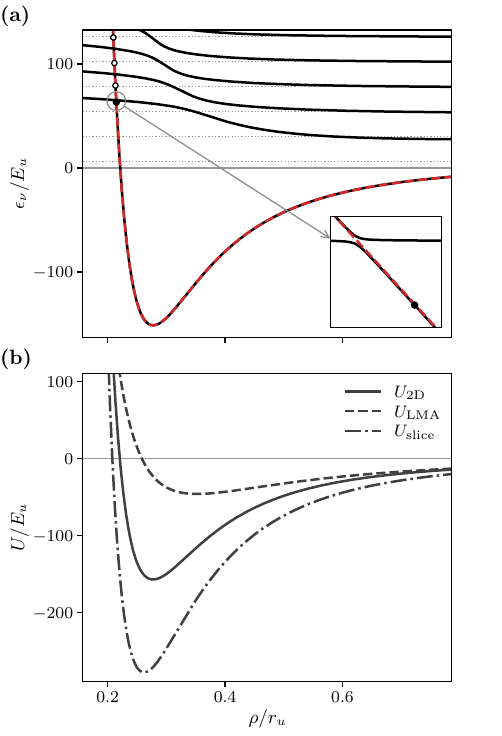}
\caption{(Color online). (a) Adiabatic energy levels $\epsilon_{\nu}(\rho\equiv|\bm{\rho}|)$ for $\nu=0,2,4,\ldots$. Horizontal dotted lines denote the non-interacting limit with $\epsilon_{\nu}^{(0)}=(\nu+1/2)\hbar\omega_z$. Red dashed line highlights $\epsilon_{0}$ and its adiabatic extension to short $\rho$ regime. Solid and open circles denote the reference state and the adiabatic state with the largest overlap with the reference state, respectively. Inset shows a magnified view near the avoided crossing between $\nu=0$ and the higher level. (b) Effective 2D potential $U_{\rm 2D}(\rho)$  (solid line), in comparison with $U_{\rm LMA}$ (dashed) and $U_{\rm slice}$ (dashed-dot). In both (a) and (b), we set $\tilde{\omega}_z=12$ and $\tilde{\Omega}=2\times10^4$. The length and energy units are respectively $r_u$ and $E_u$.}  \label{fig_potential}
\end{figure}

We have numerically solved Eq.(\ref{eq_z}) by diagonalizing the corresponding Hamiltonian on a discretized $z$ axis to obtain the eigenenergies $\epsilon_{\nu}$. In Fig.~\ref{fig_potential}(a), we show typical $\epsilon_{\nu}$ as functions of $\rho\equiv|\bm{\rho}|$ for several lowest levels with even parity along $z$, i.e., $\nu=0,2,4...$. The odd-parity levels are not shown here because different parity sectors decouple. At long distance (${\rho}\rightarrow\infty$),  these levels approach the non-interacting limit $\epsilon_{\nu}^{(0)}\sim(\nu+1/2)\hbar\omega_z$ (horizontal dotted lines in  Fig.~\ref{fig_potential}(a)), signifying the vanishing interaction effect. As $\rho$ decreases, $\epsilon_{\nu}$ starts to deviate from $\epsilon_{\nu}^{(0)}$, with the deviation most pronounced for the ground state  $\nu=0$, which determines the effective 2D potential $U_{\rm 2D}$ as in Eq.(\ref{U2d}). At short distance (small $\rho$), $\epsilon_{0}$ turns upward and undergoes avoided level crossings with higher $\nu$ levels, see the magnified plot in the inset of Fig.~\ref{fig_potential}(a). Near the crossing points, the adiabatic treatment becomes invalid because of the strong off-diagonal couplings between different $\nu$. Nevertheless, an effective 2D potential can still be identified approximately as follows. First, we choose a reference state on the $\nu=0$ curve just before the level crossing, see solid circle in Fig.~\ref{fig_potential}(a); we then search for states between neighboring crossing points that have the largest overlap with this reference state, marked by open circles in Fig.~\ref{fig_potential}(a). Connecting all these circles yields an adiabatic extension of $\epsilon_0$ to short $\rho$ regime (red dashed line in Fig.~\ref{fig_potential}(a)), which gives the full curve of $U_{\rm 2D}$ shown in Fig.~\ref{fig_potential}(b).

We remark that the above scheme for obtaining $U_{\rm 2D}$ incorporates the full interaction effect along the longitudinal direction, going well beyond the simplified treatments in $U_{\rm LMA}$ and $U_{\rm slice}$. As shown in Fig.~\ref{fig_potential}(b), $U_{\rm 2D}$ lies right between $U_{\rm LMA}$ and $U_{\rm slice}$. Physically, this is because virtual scattering into higher harmonic modes gives a negative correction to $U_{\rm LMA}$, as inferred from second-order perturbation theory; in contrast, quantum fluctuations beyond $z=0$ give a positive (zero-point) correction to $U_{\rm slice}$, as analyzed previously in the 3D case\cite{Cui1, Cui2}.   This is why our $U_{\rm 2D}$, which includes both the higher-mode contribution and the fluctuation effect beyond $z=0$, lies right between $U_{\rm LMA}$ and $U_{\rm slice}$. At long distance (large $\rho$), $U_{\rm 2D}$ matches  $U_{\rm LMA}$ well, since the interaction is sufficiently weak there to validate the lowest-mode approximation. 

\section{Comparison of binding energies}

Having identified the various 2D potentials, we can now compute the two-molecule bound states they predict and compare them with the exact results. In Figs.~\ref{fig_Emicro} and \ref{fig_Etrap}, we show how the bound states of bosonic and fermionic molecules evolve with the microwave and confinement strengths  ($\tilde{\Omega}$ and $\tilde{\omega}_z$), where we compare the exact binding energy ($E_{\rm b}$)  with those obtained from various 2D potentials, including $E_{\rm 2D}$, $E_{\rm LMA}$ and $E_{\rm slice}$ from  $U_{\rm 2D}$, $U_{\rm LMA}$ and $U_{\rm slice}$, respectively. 

\begin{widetext}

\begin{figure}[h]
\includegraphics[width=\textwidth]{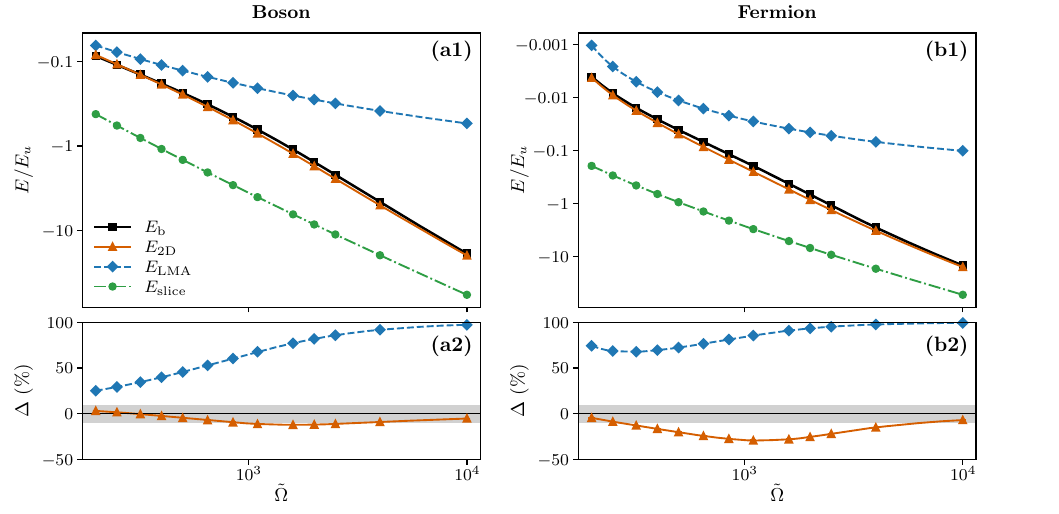}
\caption{(Color online). (a1,b1) Ground state binding energies of  bosonic (a1) and fermionic (b1) molecules as functions of $\tilde{\Omega}$ at fixed $\tilde{\omega}_z=1$. $E_{\rm b}$ is the exact solution, compared with $E_{\rm 2D}$, $E_{\rm LMA}$ and $E_{\rm slice}$  from the 2D potentials $U_{\rm 2D}$, $U_{\rm LMA}$ and $U_{\rm slice}$, respectively. (a2,b2) Relative deviations ($\Delta$, see definition in Eq.(\ref{Delta})) of various 2D binding energies  for bosonic  (a2) and fermionic (b2) systems. Here the deviation of $E_{\rm slice}$ (i.e., $\Delta_{\rm slice}$) is too large to be shown here.  The shaded area marks the region where the relative deviation is within $[-10\%,10\%]$.}  \label{fig_Emicro}
\end{figure}

\begin{figure}[t]
\includegraphics[width=\textwidth]{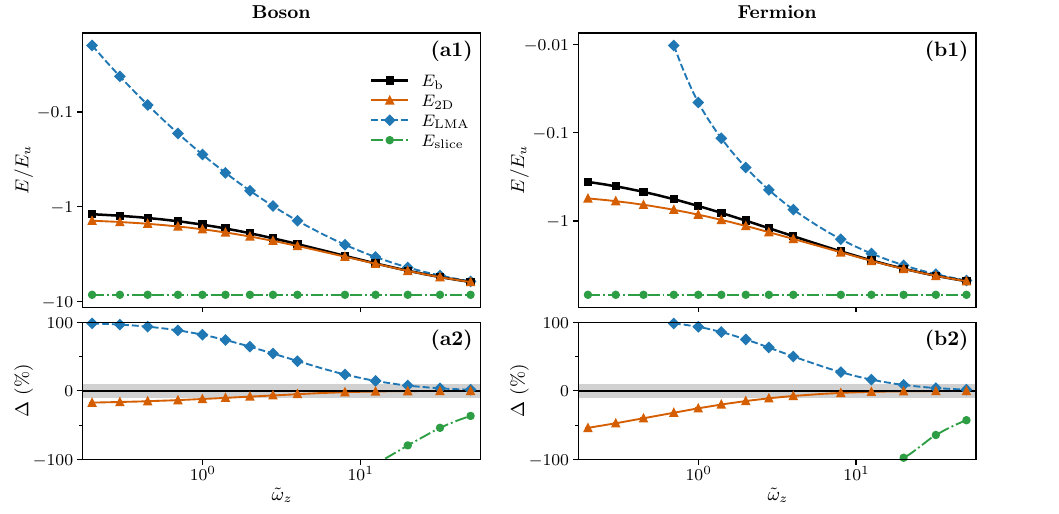}
\caption{(Color online). Similar to  Fig.~\ref{fig_Emicro},  except that the variable here is $\tilde{\omega}_z$ while  $\tilde{\Omega}=2\times10^3$ is fixed. }  \label{fig_Etrap}
\end{figure}

\end{widetext}

From Figs.~\ref{fig_Emicro}(a1,b1) and \ref{fig_Etrap}(a1,b1), we can see that the binding energies of both bosonic and fermionic molecules are  enhanced as  $\tilde{\Omega}$ or $\tilde{\omega}_z$ increases; in particular, the bosonic binding is generally deeper  than the fermionic one, owing to the presence of p-wave centrifugal barrier in the latter case. Importantly, for fixed parameters $\tilde{\Omega}$ and $\tilde{\omega}_z$, $E_{\rm 2D}$  agrees  best  with the exact $E_{\rm b}$, while $E_{\rm LMA}$ ($E_{\rm slice}$) stays well above (below) the exact results. This can be attributed to the distinct behaviors of $U_{\rm LMA}$ and $U_{\rm slice}$ as compared to $U_{\rm 2D}$, see Fig.~\ref{fig_potential}(b). Given that $U_{\rm 2D}$ has incorporated the full interaction effect along $z$, it appears as the best candidate among all 2D potentials for accurately describing the bound state formation in confined geometry.

To quantify the relative deviations of various binding energies, we define a dimensionless parameter   
\begin{equation}
\Delta\equiv\frac{E-E_{\rm b}}{|E_{\rm b}|}, \label{Delta}
\end{equation}
so that  $\Delta_{\rm 2D}$, $\Delta_{\rm LMA}$ and $\Delta_{\rm slice}$ respectively correspond to the relative deviations of  $E_{\rm 2D}$, $E_{\rm LMA}$ and $E_{\rm slice}$ from exact result. As shown in Figs.~\ref{fig_Emicro}(a2,b2) and \ref{fig_Etrap}(a2,b2), the smallest deviation always occurs for  $\Delta_{\rm 2D}$, which directly highlights the advantage of  $U_{\rm 2D}$ over the other 2D potentials in describing the system. 

Interestingly, $\tilde{\Omega}$ and $\tilde{\omega}_z$ play distinct roles in affecting the validity of various 2D potentials (reflected in $\Delta$). For instance, as  $\tilde{\Omega}$ increases (Fig.~\ref{fig_Emicro}(a2,b2)), $\Delta_{\rm LMA}$ grows while $|\Delta_{\rm 2D}|$ undergoes a non-monotonic evolution: it first increases and then decreases at large $\tilde{\Omega}$. Physically, this is because a larger $\tilde{\Omega}$  induces more higher-mode excitations that invalidate the lowest-mode approximation. In contrast, $U_{\rm 2D}$ always performs well owing to  the inclusion of higher-mode scattering; in this case, the decrease of  $|\Delta_{\rm 2D}|$ at large  $\tilde{\Omega}$ can be attributed to the stronger binding there, i.e., a larger $|E_{\rm b}|$ in the denominator of Eq.(\ref{Delta}).  
Unlike the effect of $\tilde{\Omega}$,  Fig.~\ref{fig_Etrap}(a2,b2) shows that a larger $\tilde{\omega}_z$ facilitates the validity of all 2D potentials in predicting the bound state solution, as indicated by the smaller $|\Delta_{\rm 2D}|$, $|\Delta_{\rm LMA}|$ and $|\Delta_{\rm slice}|$  as  $\tilde{\omega}_z$ increases. This is consistent with the intuitive picture that a tighter longitudinal confinement will tend to freeze the molecular scattering within the lowest harmonic mode and  also suppress quantum fluctuations beyond the $xy$ plane. 


\section{Validity diagrams of 2D potentials}

\begin{figure*}[t]
\includegraphics[width=\textwidth]{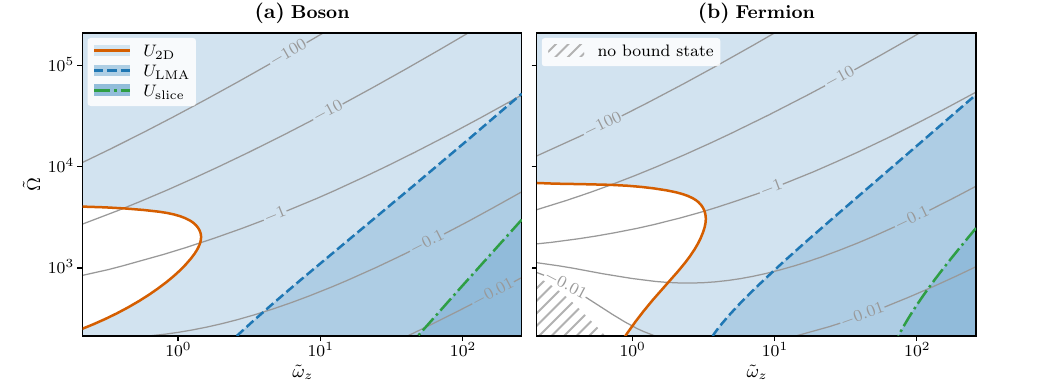}
\caption{(Color online). Validity diagrams of various 2D potentials for bosonic (a) and fermionic (b) molecules. Solid, dashed and dashed-dot curves respectively denote the boundaries at which  $|\Delta_{\rm 2D}|$, $|\Delta_{\rm LMA}|$ and $|\Delta_{\rm slice}|$ reach $10\%$. The colored area to the right  of each boundary corresponds to the validity region of according 2D potential in predicting the bound state solution, with relative energy deviation $|\Delta|<10\%$.  The  underlying lines show  contour plots of  $E_{\rm b}/(\hbar\omega_z)$. The striped area in (b) denotes the region where the bound state of fermionic molecules is forbidden.}  \label{fig_diagram}
\end{figure*}

By requiring a small relative deviation $|\Delta|$, we can identify the validity regime of various 2D potentials. Here we adopt the validity criterion $|\Delta|< 10\%$, on the basis of which we map out the validity diagrams in the ($\tilde{\omega}_z,\tilde{\Omega}$) plane, as shown in Fig.~\ref{fig_diagram}, for various 2D potentials in predicting the bound states of bosonic and fermionic molecules.

From Fig.~\ref{fig_diagram}, we can see that $U_{\rm 2D}$ has the broadest validity region, whereas $U_{\rm slice}$ has the smallest. In particular, the applicability of $U_{\rm 2D}$ can extend to much smaller $\tilde{\omega}_z$  than the other potentials, and even  to very large $\tilde{\Omega}$ regime where $U_{\rm LMA}$ and $U_{\rm slice}$ perform very poorly. Remarkably, in the large $\tilde{\Omega}$ regime where $U_{\rm 2D}$ is valid, the bound state can be quite deep, with binding energies well exceeding the confinement gap $\hbar\omega_z$, as clearly seen  from the underlying contour plots of $E_{\rm b}/(\hbar\omega_z)$ in Fig.~\ref{fig_diagram}(a,b). This   demonstrates the power of $U_{\rm 2D}$ in predicting very deep bound states. 

The validity diagrams of bosonic and fermionic systems, as respectively shown in Fig.~\ref{fig_diagram}(a) and (b),  look very similar. The only qualitative difference is   that  bound state formation in  fermionic molecules is forbidden  at small $\tilde{\Omega}$ and $\tilde{\omega}_z$, as shown by the striped area in Fig.~\ref{fig_diagram}(b). Close to the forbidden region, the shallow bound states of fermions cannot be predicted by $U_{\rm 2D}$, which may be attributed to the small adiabatic energy gap ($\sim \omega_z$) and thus an inevitable off-diagonal coupling between different $\nu$ levels. In contrast, for bosons $U_{\rm 2D}$ can predict the shallow bound states well, as shown by the colored region in Fig.~\ref{fig_diagram}(a) at small $\tilde{\Omega}$ and $\tilde{\omega}_z$. This is because in this region $\tilde{\Omega}$ is much smaller than in the fermionic case, which is thus   insufficient to couple different $\nu$ levels. This explains why the  validity of $U_{\rm 2D}$ behaves so differently for the shallow bound states in bosonic and fermionic  systems. 

Since the validity diagrams in Fig.~\ref{fig_diagram}  depend  universally on dimensionless parameters $\tilde{\Omega}$ and $\tilde{\omega}_z$, these diagrams can be directly applied to realistic bosonic and fermionic molecules with different $E_u$. Here, large (small) energy scale of $E_u$ will lead to the same large (small) energy scale of $(\Omega, \omega_z)$ at the validity boundary of Fig.~\ref{fig_diagram}. For instance, among the bosonic molecules of NaRb, NaCs and CaF, since NaCs has the smallest $E_u$, it requires the weakest strengths of microwave coupling ($\Omega$) and harmonic confinement ($\omega_z$)  to validate the  2D potentials. This also means that even a tiny $\Omega$ or $\omega_z$ will drive NaCs system to the effective 2D regime. The application of the validity diagram to other molecules can also be deduced straightforwardly.

\section{Diagonal correction to adiabatic 2D potential}

In obtaining the adiabatic 2D potential in Eq.(\ref{U2d}), we have neglected the diagonal  correction 
\begin{equation}
\Delta U(\bm{\rho})=-\frac{\hbar^2}{2\mu}\langle\chi_{0}|\nabla_{\bm{\rho}}^2\chi_{0}\rangle_z, \label{correction}
\end{equation}
which arises from the kinetic motion of ground state ($\chi_0$) in the transverse ($xy$) plane. Including this correction, the 2D potential becomes $\tilde{U}_{\rm 2D}\equiv U_{\rm 2D}+\Delta U$, and we denote the corresponding binding energy by $\tilde{E}_{\rm 2D}$. 

Before computing $\Delta U$, we note that the expression in Eq.(\ref{correction}) cannot be applied to short $\rho$ regime, where different $\nu$ levels undergo avoided level crossings and  the adiabatic treatment becomes invalid. To simplify the situation, in our numerics we have neglected the effect of $\Delta U$ in this regime,  considering that the wavefunction weight at short $\rho$  is exceedingly small owing  to the large repulsive potential there.  Specifically, we  numerically calculate $\Delta U$ up to a short-range cutoff $\rho^*$ before the first level crossing point $\rho_c$; then we decrease  $\Delta U$ smoothly to $0$ from $\rho^*$ to $\rho_c$, and keep $\Delta U=0$ for $\rho\in(0,\rho_c)$. In this way,  $\tilde{U}_{\rm 2D}$ is guaranteed as a smooth function of $\rho$.

\begin{figure*}[t]
\includegraphics[width=\textwidth]{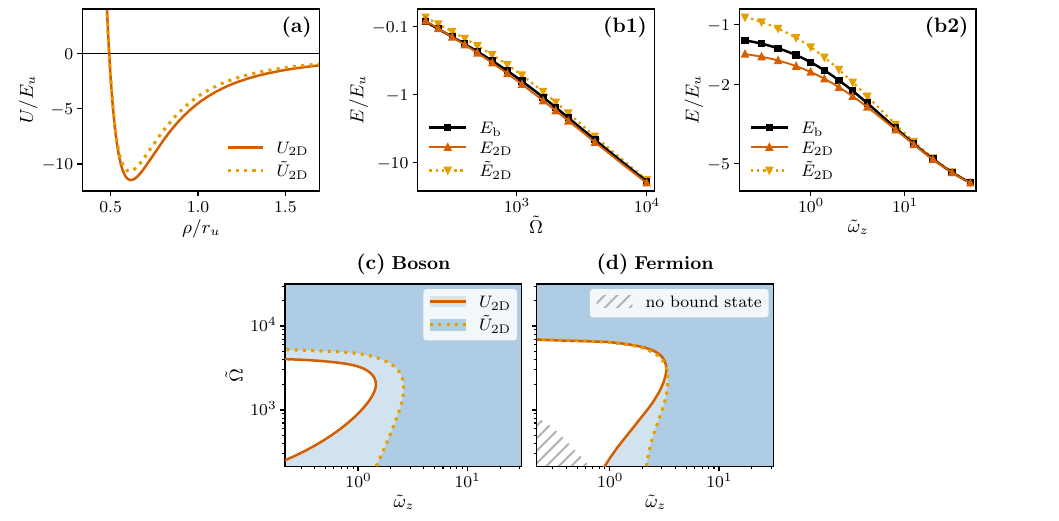}
\caption{(Color online). Effect of diagonal correction on the adiabatic 2D potential. (a) $U_{\rm 2D}$  (solid) and $\tilde{U}_{\rm 2D}\equiv U_{\rm 2D}+\Delta U$ (dotted) as functions of $\rho$ at $\tilde{\omega}_z=1$ and $\tilde{\Omega}=2\times10^3$. (b1,b2) $E_{\rm b}$, $E_{\rm 2D}$ and $\tilde{E}_{\rm 2D}$ of bosonic molecules as functions of $\tilde{\Omega}$ (b1) and $\tilde{\omega}_z$ (b2). Here we take $\tilde{\omega}_z=1$ in (b1) and $\tilde{\Omega}=2\times10^3$ in (b2). (c,d) Validity diagrams of $U_{\rm 2D}$ and $\tilde{U}_{\rm 2D}$ for bosonic (c) and fermionic (d) molecules. The length and energy units are respectively $r_u$ and $E_u$.}  \label{fig_correction}
\end{figure*}

In Fig.~\ref{fig_correction}(a), we show a typical comparison of $U_{\rm 2D}$ and $\tilde{U}_{\rm 2D}$ as functions of $\rho$. We can see that the correction $\Delta U(=\tilde{U}_{\rm 2D}-U_{\rm 2D})$ is generally positive and most pronounced at intermediate $\rho$. Therefore, $\tilde{U}_{\rm 2D}$ is less attractive than $U_{\rm 2D}$, leading to a shallower bound state than in the latter case. As shown in Fig.~\ref{fig_correction}(b1,b2), $\tilde{E}_{\rm 2D}$ always lies above the exact binding energy $E_{\rm b}$, in contrast to $E_{\rm 2D}$ that generally lies below $E_{\rm b}$. Moreover, the relative deviation of $\tilde{E}_{\rm 2D}$ from exact result is generally larger than that of $E_{\rm 2D}$. Consequently, the validity region of $\tilde{U}_{\rm 2D}$ is smaller than that of $U_{\rm 2D}$, as shown in Fig.~\ref{fig_correction}(c,d) for bosonic and fermionic systems respectively. In view of these findings, we conclude that $U_{\rm 2D}$ serves as  a better effective potential than $\tilde{U}_{\rm 2D}$ to accurately predict the bound state solutions. 


\section{Summary and outlook}

In summary, we have investigated the confinement-induced bound state and effective potential of microwave-shielded ultracold molecules under longitudinal harmonic confinement. In particular, the effective 2D potential established here incorporates the full interaction effect along the longitudinal direction and thus accurately predicts the bound-state solution over a much broader parameter regime than other simple 2D potentials. The resulting effective 2D model greatly simplifies the numerical calculations and paves the way for efficient exploration of few- and many-body physics in quasi-2D molecular systems. Practically, the universal validity diagrams in Fig.~\ref{fig_diagram} can also be directly applied to various realistic bosonic and fermionic molecules.

Beyond the quasi-2D geometry and single microwave field considered in this work, we remark that the adiabatic formulas in Eqs.(\ref{adiabatic_wf},\ref{eq_z},\ref{H2d}) are very general and can be applied to a variety of low-dimensional geometries or microwave types, such as quasi-1D geometry, finite microwave ellipticity, and double microwave shielding. Since the microwave itself can induce strong interaction anisotropy and give rise to effective dimension reduction\cite{Cui1, Cui2, Pohl}, it will be interesting to study the interplay between interaction-induced and confinement-induced dimension reductions. For instance, in our present case both the interaction-induced and confinement-induced low dimensions lie in the transverse plane. A more general situation arises, however, when the two induced low-dimensional geometries are not aligned but point along different directions. In this case, more intriguing scattering properties may emerge from the interplay of interaction anisotropy and external confinement, which warrants a careful study in the future.

\bigskip
\acknowledgments
This work is supported by National Natural Science Foundation of China (92476104, 12134015, 12525412) and Innovation Program for Quantum Science and Technology (2024ZD0300600).

\end{document}